\documentclass[12pt]{article}
\usepackage{geometry}             
\usepackage{graphicx}
\usepackage{amssymb}
\usepackage{amsmath}
\usepackage{epstopdf}
\usepackage{subfigure}
\usepackage{cite}
\usepackage{bbm}

\usepackage[hyperindex=true,
          pdfstartview=FitH,
          bookmarksnumbered=true,
          bookmarksopen=true,
          citecolor=blue,
          linkcolor=blue,
          colorlinks=true,
          pdfborder=001,
          unicode]{hyperref}

\begin{document}

\title{Circuit complexity in Proca theory}
\author{Kun Meng\thanks{email: {\it kunmeng@wfu.edu.cn}}, Meihua Deng, Yang Yang, Lianzhen Cao, Jiaqiang Zhao\\
School of Physics and Photoelectric Engineering, Weifang University, \\
  Weifang 261061, China\\
}
\date{}                             
\maketitle

\begin{abstract}
In this paper, we study circuit complexity in Proca theory with  Nielsen's approach and  Fubini-Study (FS) metric approach. We place the fields on a lattice to gain a regularized theory, and obtain the ground state by adopting proper coordinates. We calculate complexities of the ground and thermofield double (TFD) states with Nielsen's approach, complexity of the TFD state is found to grows like a logarithmic function. We quantize the Proca fields and give the approximate ground state and TFD state by acting unitary circuit operators on the associated reference states.  The circuit lengths are calculated with FS metric, the minimal lengths are given according to the associated geometric spaces. The complexity of TFD state  is found to grows linearly with time.
\end{abstract}

\section{Introduction}
In the framework of AdS/CFT, significant progress has been made in understanding the essence of spacetime from the viewpoint of information theory.
A remarkable achievement in this direction is the Ryu-Takayanagi conjecture\cite{0603001}, which states that the entanglement entropy of some region on the boundary of spacetime is proportional to the area of the codimension-2 minimal surface anchored to boundary of the region. Another significant progress in the direction is known as ER=EPR\cite{1306.0533}, it is found that for any pair of entangled black holes, there exists some kind of Einstein-Rosen bridge connecting the two black holes.

However, entanglement is found not able to access the whole holographic geometry. The interior of black hole is not captured by entanglement. Entanglement is not enough to describe the difficulty in transmitting information through Einstein-Rosen bridge\cite{1411.0690}. In order to study the interior of black hole holographically, Susskind and collaborators introduce the concept complexity in information science, and  propose two duality conjectures. The first is CV duality, which states that complexity of the  boundary state is proportional to the volume of a maximal codimension-one bulk surface that extends to the AdS boundary\cite{1406.2678,1408.2823}. The second is CA duality, which identifies complexity of boundary state as gravitational action in the Wheeler-DeWitt (WDW) pach\cite{1509.07876,1512.04993}.

As new holographic directories, CV or CA duality build new connections between bulk gravity theory and boundary CFT. The gravity part has been studied a lot\cite{1512.04993,1703.06297,1612.03627,1606.08307,1806.06216,1804.07410,1805.07262,1806.10312,1808.09917,1808.00067,2107.08608,1810.02208}, including various gravity systems and various types of black holes. On the part of boundary quantum field theory (QFT), it is natural to ask what does complexity mean for boundary QFT states. Complexity, as a concept of information theory, denotes the minimal number of basic logic gates that are needed to implement a given task. Thus it is natural to define complexity for QFT states  through searching for the optimal circuit that is shortest in states space. Among the approaches to define complexity, the Nielsen's approach, which is inspired by Nielsen's geometric idea\cite{0502070}, gives complexity through searching for a path that minimize some physically reasonable cost function. With this approach, complexities of the ground and TFD states in free scalar and spinor field theories were studied in Refs. \cite{1707.08570,1801.07620,1803.10638,1810.05151,1812.00193,1910.08806,1909.10557,2004.00344}. It is found that complexity obtained with Nielsen's approach gives the volume law scaling which is consistent with holography\cite{1907.08223}. The FS metric approach measure circuits length by the FS line element, and identify complexity as the length of the shortest path connecting reference and target states. Complexities of the ground and TFD states in free scalar and spinor field theories have also been studied with this approach \cite{1707.08582,1710.00600,1801.07620,1902.01912,2103.06920}. Path-integral approach is another attempt to define complexity, with this approach complexities of some CFT states were given in Refs.\cite{1703.00456,1706.07056,2011.08188,1807.04422}. Circuit complexities obtained with above approaches have been used to study phase transitions of some condensed matter systems recently\cite{1811.05985,1902.10720,1906.11279,2106.12648}.

However, it is noted that, for both Nielsen's and FS metric approaches, present investigations mainly involve Gaussian ground state or TFD state in free QFTs. Although there have been a few attempts to study circuit complexities in self-interacting scalar field theories\cite{1808.03105,2109.09759}, calculable complexities of general non-Gaussian states and states in interacting theories are far from complete.
Even for vector fields, which are important constituents of interacting theories and play the role of medium of basic interactions, complexity of the associate QFT state has not been studied\footnote{After the first version of this paper presented in arXiv, Ref.\cite{2108.08208}, which has some overlap with this paper, presented in arXiv immediately.}. In this paper we will study circuit complexity in a vector field theory---the Proca theory---with both Nielsen's  and FS metric approaches. We wish our attempts will serve as a catalyst for further study of complexity in interacting theories.
This paper is organized as, in section \ref{section2}, we study complexity in Proca theory with Nielsen's approach. We study complexity of ground state in subsection \ref{CgroundN} and that of TFD state in subsection \ref{CTFDN}. In section \ref{section3}, we study complexity in Proca theory with FS metric approach. Complexities of ground state and TFD state are studies respectively in subsections \ref{CgroundFS} and \ref{CTFDFS}. We summarize our results in the last section.

\section{Nielsen's approach\label{section2}}
In this section we study complexity with Nielsen's approach. We will consider the ground state and TFD state.
\subsection{complexity of ground state\label{CgroundN}}
Let's first review the Nielsen's approach. Suppose the target and reference states are connected by a unitary operator $|\psi_T\rangle=U|\psi_R\rangle$, where $U$ is synthesized by path-ordered gate operators
\begin{align}
U=\overleftarrow{\mathcal{P}}\exp\int_0^1ds Y^I(s)\mathcal{O}_I,
\end{align}
where $\mathcal{O}_I$ form the set of elementary gate operators of our problem, the coefficients $Y^I(s)$ may be turned on or off to specify a trajectory, thus $Y^I(s)$ can be seen as the components of velocity vector tangent to the trajectory.

The essence of Nielsen's approach is to search for a path that minimize some cost function. The physically reasonable cost functions are given by\cite{0502070}
\begin{align}
F_1(U,Y)&=\sum_I|Y^I|,\;\;\;\;\;\;\;\;\;\;\;F_p(U,Y)=\sum_Ip_I|Y^I|,\nonumber\\
F_2(U,Y)&=\sqrt{\sum_I(Y^I)^2},\;\;\;\;\;\;F_q(U,Y)=\sqrt{\sum_Iq_I(Y^I)^2}.
\end{align}
In this paper, we select $F_2$ to measure complexity.

Usually it is complicated to find the minimal $F_2$, a useful tool that makes the task easier is the covariance matrix approach, now we briefly review this approach. For a bosonic system with canonical coordinates $\xi^a$, the two-point function is given by
\begin{align}
\langle\psi|\xi^a\xi^b|\psi\rangle=\frac{1}{2}(G^{a,b}+i\Omega^{a,b}),
\end{align}
where $G^{a,b}=G^{(a,b)}$ denotes the symmetric part of the two-point function and $\Omega^{a,b}=\Omega^{[a,b]}$ denotes the antisymmetric part. For bosonic degrees of freedom, the antisymmetric part is completed fixed by the canonical commutation relations to
\begin{align}
\Omega^{a,b}=\left(
                     \begin{array}{cc}
                       0 & \mathbbm{1} \\
                       -\mathbbm{1} & 0 \\
                     \end{array}
                   \right),
\end{align}
therefore the state $|\psi\rangle$ is completely characterized by the covariance matrix
\begin{align}
G^{a,b}=\langle\psi|\xi^a\xi^b+\xi^b\xi^a|\psi\rangle.
\end{align}

The most general quadratic operator is given by
\begin{align}
\hat{K}=\frac{1}{2}\xi^ak_{a,b}\xi^b=\frac{1}{2}\xi k\xi^{\intercal},
\end{align}
where  $k_{a,b}$ is symmetric in the two subscripts since $\xi^a\xi^b$ is symmetric. The quadratic operator $\hat{K}$ generates a unitary connecting two Gaussian states
\begin{align}
\hat{U}=e^{-i\sigma\hat{K}},\;\;\;\;\;\;|G_\sigma\rangle=\hat{U}|G_0\rangle.
\end{align}
To find the relation between the covariance matrices of the states $|G_0\rangle$ and $|G_\sigma\rangle$, one acts the unitary  $\hat{U}$ on the canonical coordinate $\xi^a$
\begin{align}
\hat{U}^{\dag}\xi^a\hat{U}=\sum_{n=0}^\infty\frac{\sigma^n}{n!}[i\hat{K},\xi^a]_{(n)},\label{UxiU}
\end{align}
where $[i\hat{K},\xi^a]_{(n)}$ is defined recursively as $[i\hat{K},\xi^a]_{(n)}=[i\hat{K},[i\hat{K},\xi^a]_{(n-1)}]$, and $[i\hat{K},\xi^a]_{(0)}=\xi^a$. Considering the commutation relation $[\xi^a,\xi^b]=i\Omega^{a,b}$, it is straightforward to obtain
\begin{align}
[i\hat{K},\xi^a]=\Omega^{a,b}k_{b,c}\xi^c=K^a_{\;\;b}\xi^b,\label{mgenerator}
\end{align}
here the matrix generator $K^a_{\;\;b}=\Omega^{a,b}k_{b,c}$ associated to operator $\hat{K}$ is introduced. For a bosonic system with $N$ degrees of freedom, one can check that $K\in\mathfrak{sp}(2N,\mathbb{R})$ and $U(\sigma)=e^{\sigma K}$ preserves the symplectic structure
\begin{align}
U(\sigma)\Omega U^{\intercal}(\sigma)=\Omega,
\end{align}
i.e., $U(\sigma)=e^{\sigma K}$ is element of the symplectic group $\textrm{Sp}(2N,\mathbb{R})$. Combining (\ref{UxiU}) and (\ref{mgenerator}) one has
\begin{align}
\hat{U}^{\dag}\xi^a\hat{U}=U(\sigma)^a_{\;\;b}\xi^b.
\end{align}
Now the relation between the covariance matrices of $|G_0\rangle$ and $|G_\sigma\rangle$ can be given by
\begin{align}
G_\sigma^{a,b}&=\langle G_\sigma|\xi^a\xi^b+\xi^b\xi^a|G_\sigma\rangle\nonumber\\
&=\langle G_0|e^{i\sigma\hat{K}}(\xi^a\xi^b+\xi^b\xi^a)e^{-i\sigma\hat{K}}|G_0\rangle\nonumber\\
&=U(\sigma)^a_{\;\;c}G_0^{c,d}U(\sigma)^b_{\;\;d}.
\end{align}

It has been proven in Ref.\cite{1810.05151} that the path
\begin{align}
\gamma(\sigma)=e^{\sigma K}
\end{align}
is the shortest one that connects $\mathbbm{1}$ and an arbitrary point $U=e^{K}$ in the group space of $\mathrm{Sp}(2N,\mathbb{R})$. The covariance matrices of reference and target states are connected by
\begin{align}
G_T=e^K G_R e^{K^\intercal}.
\end{align}
The relation $G_T=\sqrt{G_Tg_R} G_R (\sqrt{G_Tg_R})^\intercal$ can be checked to be satisfied, with $G_Rg_R=\mathbbm{1}$, thus one has
\begin{align}
K=\log U=\frac{1}{2}\log G_Tg_R.\label{K}
\end{align}

With the Nielsen's approach given above, let's now calculate complexity in Proca theory. The Lagrangian density of the theory is given by
\begin{align}
\mathcal{L}=-\frac{1}{4}F_{\mu\nu}F^{\mu\nu}+\frac{1}{2}m^2A_\mu A^\mu.\label{lag}
\end{align}
In terms of the conjugate momenta
\begin{align}
\pi^0=\frac{\partial\mathcal{L}}{\partial \dot{A}_0}=0,\;\;\;\;\;\;\;\;\;\;\;\;\;\;\;\;\;\;\;
\pi^i=\frac{\partial\mathcal{L}}{\partial \dot{A}_i}=-\dot{A}^i,\label{momentum}
\end{align}
one obtains the Hamiltonian of Proca theory
\begin{align}
H&=\int d^{d-1}x(\pi^i\dot{A}_i-\mathcal{L}),\nonumber\\
&=\frac{1}{2}\int d^{d-1}x\left[(\pi^i)^2+(\partial_iA_j-\partial_jA_i)\partial_iA_j+m^2(A_i)^2\right].\label{Ham0}
\end{align}
Note that, due to the primary constraint $\pi^0=0$, $A_0$ has zero Poisson bracket with any physical quantity. Thus $A_0$ and $\pi^0$ are not of physically interest and can be eliminated from theory. After eliminating $A_0$ and $\pi^0$ one obtains the Hamiltonian (\ref{Ham0}) which still retains all the degrees of freedom which are physically interest. It is easy to check the Hamiltonian (\ref{Ham0}) gives rise to the equations of motion that are consistent with the Lagrangian equations.

To obtain a regularized theory, we place the theory on a $d-1$-lattice, then the Hamiltonian becomes
\begin{align}
H&=\frac{1}{2}\sum_{\vec{n}}\left\{\sum_{i}\frac{(P^i(\vec{n}))^2}{\delta^{d-1}}+\delta^{d-1}\sum_{j\neq i}\sum_{i}
\left[\frac{1}{\delta^2}(A_j(\vec{n})-A_j(\vec{n}-\hat{x}_i))^2,\right.\right.\nonumber\\
&\left.\left.-\frac{1}{\delta^2}(A_i(\vec{n})-A_i(\vec{n}-\hat{x}_j))(A_j(\vec{n})-A_j(\vec{n}-\hat{x}_i))+m^2\sum_i(A_i(\vec{n}))^2\right]\right\},
\end{align}
where $\delta$ is the lattice spacing, $\vec{n}$ is the position of any lattice site, $\hat{x}_i$ is unit vector along the $i$-direction, and the momentum $P^i(\vec{n})$ is introduced as $P^i(\vec{n})=\pi^i(\vec{n})\delta^{d-1}$. We make the redefinitions $X_i(\vec{n})=A_i(\vec{n})\delta^{d/2},P^i(\vec{n})=p^i(\vec{n})/\delta^{d/2}, M=1/\delta, \omega=m, \Omega=1/\delta$, and then obtain the Hamiltonian of coupled harmonic oscillators
\begin{align}
H=\sum_{\vec{n}}\left[\sum_{i}\frac{(p^i(\vec{n}))^2}{2M}+\frac{1}{2}M\omega^2\sum_{i}X_i^2(\vec{n})+\frac{1}{2}M\Omega^2\sum_{j\neq i}\sum_{i}\left(X_i(\vec{n})-X_i(\vec{n}-\hat{x}_j)\right)^2\right.\nonumber\\
\left.-\frac{1}{2}M\Omega^2\sum_{j\neq i}\sum_{i}\left(X_i(\vec{n})-X_i(\vec{n}-\hat{x}_j)\right)
\left(X_j(\vec{n})-X_j(\vec{n}-\hat{x}_i)\right)\right]
\end{align}
In the following of the paper, we will only consider the spacetime with two spatial dimensions for simplicity. We denote one spatial component of the field $X$  as $x$ and the other spatial component of $X$ as $y$, then the Hamiltonian becomes
\begin{align}
H=\sum_{a,b=0}^{N-1}\left[\frac{(p^x_{a,b})^2}{2M}+\frac{(p^y_{a,b})^2}{2M}+\frac{1}{2}M\omega^2x_{a,b}^2+\frac{1}{2}M\omega^2y_{a,b}^2
+\frac{1}{2}M\Omega^2(x_{a,b}-x_{a,b-1})^2\right.\nonumber\\
\left.+\frac{1}{2}M\Omega^2(y_{a,b}-y_{a-1,b})^2-M\Omega^2(x_{a,b}-x_{a,b-1})(y_{a,b}-y_{a-1,b})\right],\label{2dcoupled}
\end{align}
here we use the subscript $a,b$ to denote the location of a lattice site. For any spatial dimension we insert $N$ lattice sites, and add the periodic boundary conditions $x_{a,b}=x_{N+a,N+b}$ and $y_{a,b}=y_{N+a,N+b}$, this form a torus of $N^2$ harmonic oscillators. In order to find the ground state of the system (\ref{2dcoupled}), we try to write (\ref{2dcoupled}) into the form of decoupled harmonic oscillators. First we make the coordinate transformation
\begin{align}
\tilde{x}_{k_1,k_2}=\frac{1}{N}\sum_{a,b=0}^{N-1}\exp\left(\frac{-i2\pi k_1a}{N}\right)\exp\left(\frac{-i2\pi k_2b}{N}\right)x_{a,b},\nonumber\\
\tilde{y}_{k_1,k_2}=\frac{1}{N}\sum_{a,b=0}^{N-1}\exp\left(\frac{-i2\pi k_1a}{N}\right)\exp\left(\frac{-i2\pi k_2b}{N}\right)y_{a,b},\label{CT1}
\end{align}
where  $k_1, k_2\in[0, N-1]$.  With the new coordinates $\tilde{x}_{k_1,k_2}, \tilde{y}_{k_1,k_2}$ the Hamiltonian can be rewritten as
\begin{align}
H&=\sum_{k_1,k_2=0}^{N-1}\left[\frac{|\tilde{p}^x_{k_1,k_2}|^2}{2M}+\frac{|\tilde{p}^y_{k_1,k_2}|^2}{2M}
+\frac{1}{2}M\left(\omega^2+4\Omega^2\sin^2\frac{\pi k_1}{N}\right)|\tilde{x}_{k_1,k_2}|^2\right.\nonumber\\
&\left.+\frac{1}{2}M\left(\omega^2+4\Omega^2\sin^2\frac{\pi k_1}{N}\right)|\tilde{y}_{k_1,k_2}|^2
-M\Omega^2(1-e^{\frac{i2\pi k_1}{N}})\tilde{x}_{k_1,k_2}(1-e^{\frac{-i2\pi k_2}{N}})\tilde{y}_{k_1,k_2}^{\dag}\right].
\end{align}
It is easy to see that unlike $x_{a,b}$ and $y_{a,b}$, the coordinates $\tilde{x}_{k_1,k_2}$ and $\tilde{y}_{k_1,k_2}$ are complex. In the above derivation  the relations $\tilde{x}^{\dagger}_{k_1,k_2}=\tilde{x}_{N-k_1,N-k_2}$, $\tilde{x}_{N-k_1,N-k_2}=\tilde{x}_{-k_1,-k_2}$, $\tilde{y}^{\dagger}_{k_1,k_2}=\tilde{y}_{N-k_1,N-k_2}$  and $\tilde{y}_{N-k_1,N-k_2}=\tilde{y}_{-k_1,-k_2}$ have been used. These relations indicate the positive and negative modes are mixed, the two complex degrees of freedom labeled by ($\tilde{x}_{k_1,k_2}, \tilde{p}^x_{k_1,k_2}$) and ($\tilde{x}_{-k_1,-k_2}, \tilde{p}^x_{-k_1,-k_2}$) actually contain only two real degrees of freedom, similarly for ($\tilde{y}_{k_1,k_2}, \tilde{p}^y_{k_1,k_2}$) and ($\tilde{y}_{-k_1,-k_2}, \tilde{p}^y_{-k_1,-k_2}$). Therefore, using the complex coordinates $\tilde{x}_{k_1,k_2}$ and $\tilde{y}_{k_1,k_2}$ does not bring extra degrees of freedom. Next we make a further coordinate transformation
\begin{align}
\bar{x}_{k_1,k_2}&=\frac{\sin\frac{k_1\pi}{N}}{\sqrt{\sin^2\frac{k_1\pi}{N}+\sin^2\frac{k_2\pi}{N}}}\tilde{x}_{k_1,k_2}+
\frac{\sin\frac{k_2\pi}{N}}{\sqrt{\sin^2\frac{k_1\pi}{N}+\sin^2\frac{k_2\pi}{N}}}\tilde{y}_{k_1,k_2},\nonumber\\
\bar{y}_{k_1,k_2}&=-\frac{e^{\frac{i(k_1-k_2)\pi}{N}}\sin\frac{k_2\pi}{N}}{\sqrt{\sin^2\frac{k_1\pi}{N}+\sin^2\frac{k_2\pi}{N}}}\tilde{x}_{k_1,k_2}+
\frac{e^{\frac{i(k_1-k_2)\pi}{N}}\sin\frac{k_1\pi}{N}}{\sqrt{\sin^2\frac{k_1\pi}{N}+\sin^2\frac{k_2\pi}{N}}}\tilde{y}_{k_1,k_2},
\end{align}
with which the Hamiltonian can be recast in the form
\begin{align}
H=\sum_{k_1,k_2=0}^{N-1}\left(\frac{|\bar{p}^x_{k_1,k_2}|^2}{2M}+\frac{|\bar{p}^y_{k_1,k_2}|^2}{2M}
+\frac{1}{2}M\bar{\omega}_{k_1,k_2}^2|\bar{x}_{k_1,k_2}|^2+\frac{1}{2}M \omega^2|\bar{y}_{k_1,k_2}|^2\right),\label{Hdiag}
\end{align}
where $\bar{\omega}_{k_1,k_2}^2\equiv\omega^2+4\Omega^2(\sin^2\frac{\pi k_1}{N}+\sin^2\frac{\pi k_2}{N})$. Now, the problem has been reduced to the one of decoupled harmonic oscillators, which enables us easily to give the ground state of the system as
\begin{align}
\psi_0=\prod_{k_1,k_2=0}^{N-1}\left(\frac{M\bar{\omega}_{k_1,k_2}}{\pi}\right)^{\frac{1}{4}}
\exp\left(-\frac{1}{2}M\bar{\omega}_{k_1,k_2}|\bar{x}_{k_1,k_2}|^2\right)
\cdot \left(\frac{M\omega}{\pi}\right)^{\frac{1}{4}}\exp\left(-\frac{1}{2}M\omega|\bar{y}_{k_1,k_2}|^2\right),\label{groundN}
\end{align}
this state is selected as the target state in this subsection.

To calculate complexity, we need to choose a reference state, which is given by
\begin{align}
\psi_R=\prod_{k_1,k_2=0}^{N-1}\left(\frac{M\mu}{\pi}\right)^{\frac{1}{2}}
\exp\left[-\frac{1}{2}M\mu(|\bar{x}_{k_1,k_2}|^2+|\bar{y}_{k_1,k_2}|^2)\right],\label{referenceN}
\end{align}
here the fixed parameter $\mu$ is used to characterize the reference state. For latter convenience, it is useful to introduce the parameter $\omega_g$ and define the dimensionless position and momentum
$
\hat{x}\equiv\omega_g\bar{x},\;\hat{y}\equiv\omega_g\bar{y},\;\hat{p}\equiv\frac{\bar{p}}{\omega_g}$,
with which the reference and ground states can be rewritten as
\begin{align}
\psi_R&=\prod_{k_1,k_2=0}^{N-1}\sqrt{\frac{\lambda_R}{\pi}}\exp\left[-\frac{\lambda_R}{2}(|\hat{x}_{k_1,k_2}|^2+|\hat{y}_{k_1,k_2}|^2)\right],\nonumber\\
\psi_0&=\prod_{k_1,k_2=0}^{N-1}\left(\frac{\hat{\lambda}_{k_1,k_2}}{\pi}\right)^{\frac{1}{4}}
\exp\left(-\frac{1}{2}\hat{\lambda}_{k_1,k_2}|\hat{x}_{k_1,k_2}|^2\right)
\cdot \left(\frac{\lambda}{\pi}\right)^{\frac{1}{4}}\exp\left(-\frac{1}{2}\lambda|\hat{y}_{k_1,k_2}|^2\right),\label{groundN2}
\end{align}
where $\lambda_R=M\mu/\omega_g^2$, $\hat{\lambda}_{k_1,k_2}=M\bar{\omega}_{k_1,k_2}/\omega_g^2$ and $\lambda=M\omega/\omega_g^2$.

Straightforward calculations give the covariance matrices of the target and reference states
\begin{align}
G_T=\left(
      \begin{array}{cccccccccc}
        \frac{1}{\hat{\lambda}_{1,1}} & \; & \; & \; & \; & \; & \; & \; & \; & \; \\
        \; & \frac{1}{\hat{\lambda}_{1,2}} & \; & \; & \; & \; & \; & \; & \; & \; \\
        \; & \; & \ddots & \; & \; & \; & \; &{\Huge0} & \; & \; \\
        \; & \;& \; & \frac{1}{\lambda} & \; & \; & \; & \; & \; & \; \\
        \;& \; & \; & \; & \frac{1}{\lambda} & \; & \; & \; & \; & \; \\
        \; & \; & \; & \; & \; & \hat{\lambda}_{1,1} & \; & \; & \; & \; \\
        \; & {\Huge0}& \; & \; & \; & \; & \hat{\lambda}_{1,2} & \; & \; & \; \\
        \; &\; & \; & \; & \; & \; & \; & \ddots & \; & \; \\
        \; & \; & \; & \; & \; & \; & \; & \; & \lambda & \; \\
        \; & \; & \; & \; & \; & \; & \; & \; & \; & \lambda \\
      \end{array}
    \right),
    G_R=\left(
      \begin{array}{cccccc}
        \frac{1}{\lambda_R} & \; & \; & \; & \; & \; \\
        \; & \frac{1}{\lambda_R} & \; & \; & 0 & \; \\
        \; & \; & \ddots & \; & \; & \; \\
        \; & \; & \; & \lambda_R & \; & \; \\
        \; & 0 & \; & \; & \lambda_R & \; \\
        \; & \; & \; & \; & \; & \ddots \\
      \end{array}
    \right).
\end{align}
We choose $\lambda_R=1$ for simplicity, then the covariance matrix of reference state becomes identity matrix.
In terms of the relative covariance matrix $\Delta\equiv G_Tg_R$, complexity of the ground state is given by
\begin{align}
\mathcal{C}_2&=\big(\sum_I(Y^I)^2\big)^{1/2}=\frac{1}{2\sqrt{2}}\sqrt{\textmd{Tr}[(\log\Delta)^2]}\nonumber\\
&=\frac{1}{2}\sqrt{\sum_{k_1,k_2=0}^{N-1}\big[(\log\frac{\bar{\omega}_{k_1,k_2}}{\mu})^2+(\log\frac{\omega}{\mu})^2\big]}.\label{complexNG}
\end{align}
One sees from this result that, as $\bar{\omega}_{k_1,k_2}$ or $\omega$ increases while keep $\mu$ fixed complexity increases. That is to say, evolving a reference state to a target state with higher energy needs more quantum operators. Similar result obtained for free scalar field theory  in Ref.\cite{1707.08570}. Comparing our Eq.(\ref{complexNG}) with Eq.(4.32) in Ref.\cite{1707.08570}, one sees that the term $(\log\frac{\omega}{\mu})^2$ only present in Eq.(\ref{complexNG}) here. The term  $(\log\frac{\omega}{\mu})^2$ arises from the $y$ part of the Hamiltonian (\ref{Hdiag}) while the term $(\log\frac{\bar{\omega}_{k_1,k_2}}{\mu})^2$ arises from the $x$ part of (\ref{Hdiag}). We will see below the contribution of the $y$ part to complexity plays an important role in determining the profile of the time evolution curve of TFD-state complexity.

\subsection{complexity of \textrm{TFD} state\label{CTFDN}}
TFD state is the entanglement state of two CFTs defined respectively on the left/right boundary of the Penrose diagram of an AdS black hole
\begin{align}
|\textrm{TFD}(t_L,t_R)\rangle=\frac{1}{\sqrt{Z_\beta}}\sum_n e^{-\beta E_n/2}e^{-iE_n(t_L+t_R)}|n\rangle_L|n\rangle_R,
\end{align}
where $|n\rangle_{L,R}$ and $t_{L,R}$ are the energy eigenstates and times of the left/right CFT, and $Z_\beta$ is the canonical partition function with inverse temperature $\beta$. For a pair of entanglement simple harmonic oscillators which
are denoted by subscripts $L$ and $R$ respectively, if $t_{L,R}$ are set to  $t_L=t_R=t/2$, one has
\begin{align}
|\textrm{TFD}(t)\rangle=\left(1-e^{-\beta\omega}\right)^{1/2}\sum_{n=0}^\infty e^{-n\beta\omega/2}e^{-i(n+\frac{1}{2})\omega \label{TFDt} t}|n\rangle_L|n\rangle_R.
\end{align}
The above time-dependent \textrm{TFD} state can be rewritten with the creation and annihilation operators as
\begin{align}
|\textrm{TFD}(t)\rangle=\exp(z a_L^{\dagger}a_R^{\dagger}-z^{*}a_La_R)|0\rangle_L|0\rangle_R,
\end{align}
where $z=\alpha e^{-i\omega t}$, $\tanh\alpha\equiv\exp(-\beta\omega/2)$ and $|0\rangle_{L,R}$ are the left/right ground states.
For latter convenience, we choose the canonical coordinates
$
q_{\pm}=\frac{1}{\sqrt{2}}(q_L\pm q_R),\;p_{\pm}=\frac{1}{\sqrt{2}}(p_L\pm p_R),\label{canonicalC}
$
with $q_{L,R}$ and $p_{L,R}$ being the left/right dimensionless  position and momentum. Utilizing the canonical coordinates, and considering the relations between canonical coordinates and creation-annihilation operators, the \textrm{TFD} state can be rewritten in a factorized form
\begin{align}
|\textrm{TFD}(t)\rangle=\exp[-i\alpha \hat{\mathcal{O}}_{+}(t)]|0\rangle_{+}\otimes \exp[i\alpha \hat{\mathcal{O}}_{-}(t)]|0\rangle_{-},\label{factorTFD}
\end{align}
with $\hat{\mathcal{O}}_{\pm}(t)=\frac{1}{2}\cos(\omega t)(q_{\pm}p_{\pm}+p_{\pm}q_{\pm})+\frac{1}{2}\sin(\omega t)(\lambda q_{\pm}^2-\frac{1}{\lambda}p_{\pm}^2)$.

Now, let's calculate covariance matrix of the state (\ref{factorTFD}). Since the $\pm$ modes decouple, we could calculate the covariance matrices of the $\pm$ mode states separately. From the expression of $\hat{\mathcal{O}}_{\pm}(t)$ it is easy to read out $k_{11}=\lambda \sin(\omega t), k_{12}=k_{21}=\cos(\omega t), k_{22}=-\frac{1}{\lambda}\sin(\omega t)$. Considering $K^a_{\;\;b}=\Omega^{ac}k_{cb}$, one has
\begin{align}
K=\left(
    \begin{array}{cc}
      \cos(\omega t) & -\frac{1}{\lambda}\sin(\omega t) \\
      -\lambda \sin(\omega t) & -\cos(\omega t) \\
    \end{array}
  \right),
\end{align}
which generates the matrix $U^{+}=e^{\alpha K}$ corresponding to the operator $\exp[-i\alpha \hat{\mathcal{O}}_{+}(t)]$.
With $U^{+}$, the $+$ mode  TFD state can be achieved by
\begin{align}
G^{+}_{\textrm{TFD}}&=U^{+}G_0(U^{+})^{T}\nonumber\\
&=\left(
                                         \begin{array}{cc}
                                           \frac{1}{\lambda}[\cosh(2\alpha)+\sinh(2\alpha)\cos(\omega t)] & -\sinh(2\alpha)\sin(\omega t) \\
                                           -\sinh(2\alpha)\sin(\omega t) & \lambda[\cosh(2\alpha)-\sinh(2\alpha)\cos(\omega t)] \\
                                         \end{array}
                                       \right),\label{TFDp}
\end{align}
where $G_0$ is  covariance matrix of the ground state. For the $-$ mode, one just needs to make the replacement $\alpha\mapsto-\alpha$, i.e., one has $U^{-}=U^{+}(\alpha\rightarrow-\alpha)$ and  $G^{-}_{\textrm{TFD}}=G^{+}_{\textrm{TFD}}(\alpha\rightarrow-\alpha)$.

To apply the results of giving the shortest path found in Ref.\cite{1707.08570},
we use the  generally parameterized element of $\textrm{Sp}(2,\mathbb{R})$
\begin{align}
U(\rho,\theta,\tau)=\left(
                      \begin{array}{cc}
                        \cos\tau\cosh\rho-\sin\theta\sinh\rho & -\sin\tau\cosh\rho+\cos\theta\sinh\rho \\
                        \sin\tau\cosh\rho+\cos\theta\sinh\rho  & \cos\tau\cosh\rho+\sin\theta\sinh\rho \\
                      \end{array}
                    \right),\label{generalU}
\end{align}
to achieve the target state. The cost function $F_2$ is associated to the following right-invariant metric
\begin{align}
\textrm{d}s^2&=\frac{1}{2}\textrm{Tr}(\textrm{d}UU^{-1}(\textrm{d}UU^{-1})^\intercal)\nonumber\\
&=\textrm{d}\rho^2+\cosh(2\rho)\cosh^2\rho\textrm{d}\tau^2+\cosh(2\rho)\sinh^2\rho\textrm{d}\theta^2
-\sinh^2(2\rho)\textrm{d}\tau\textrm{d}\theta.\label{metric}
\end{align}
The Killing vectors associated to the metric (\ref{metric}) can be given through solving the Killing equations. For every Killing vector there exist a corresponding conserved momentum. Based on the Killing vectors and conserved momenta, it is found in Ref.\cite{1707.08570} that the geodesic in the space  (\ref{metric}) is given by
\begin{align}
\rho(\sigma)=\rho_1\sigma,\;\;\;\;\;\;\theta(\sigma)=\theta_0=\theta_1,\;\;\;\;\;\;\tau(\sigma)=0.
\end{align}

In terms of the $\mathfrak{sp}(2,\mathbb{R})$ generators
\begin{align}
W=\left(
    \begin{array}{cc}
      1 & 0 \\
      0 & -1 \\
    \end{array}
  \right),\;\;\;\;\;\;V=\left(
                          \begin{array}{cc}
                            0 & 0 \\
                            -\sqrt{2} & 0 \\
                          \end{array}
                        \right),\;\;\;\;\;\;Z=\left(
                                                \begin{array}{cc}
                                                  0 & \sqrt{2} \\
                                                  0 & 0 \\
                                                \end{array}
                                              \right),
\end{align}
the circuit (\ref{generalU}) can be rewritten as
\begin{align}
U(\sigma)=\exp\left[-\rho_1\sin\theta_1\sigma W+\frac{\rho_1}{\sqrt{2}}\cos\theta_1\sigma(Z-V)\right],\label{Upositive}
\end{align}
from which the complexity can be read out as
\begin{align}
\mathcal{C}_2=\sqrt{\sum_I(Y^I)^2}=\rho_{1}
\end{align}

By acting the $\textrm{Sp}(2,\mathbb{R})$ element (\ref{generalU}) on the covariance matrix of reference state one obtains
\begin{align}
G_T&=U(\rho,\theta,\tau)G_R U^{T}(\rho,\theta,\tau)\nonumber\\
&=\left(
    \begin{array}{cc}
      \cosh(2\rho)-\sin(\theta+\tau)\sinh(2\rho) & \cos(\theta+\tau)\sinh(2\rho) \\
      \cos(\theta+\tau)\sinh(2\rho) & \cosh(2\rho)+\sin(\theta+\tau)\sinh(2\rho) \\
    \end{array}
  \right).\label{targetTFD}
\end{align}
Note that $G_R=\mathbbm{1}$ for the choice $\lambda_R=1$.
We choose the time-dependent \textrm{TFD} state (\ref{TFDp}) as the target state, by comparing (\ref{TFDp}) with (\ref{targetTFD}) yields
\begin{align}
\cosh(2\rho_1)=\frac{1+\lambda^2}{2\lambda}\cosh(2\alpha)+\frac{1-\lambda^2}{2\lambda}\sinh(2\alpha)\cos(\omega t).
\end{align}
Combining the $\pm$ modes and considering the relation $\cosh^{-1}x=\log(x+\sqrt{x^2-1})$, complexity of the TFD state (\ref{TFDt}) is given by
\begin{align}
\mathcal{C}_2=\frac{1}{2}\left[\log^2\left(f^{(+)}+\sqrt{(f^{(+)})^2-1}\right)
+\log^2\left(f^{(-)}+\sqrt{(f^{(-)})^2-1}\right)\right]^{1/2},
\end{align}
where $f^{(+)}=\cosh(2\rho_1)=\frac{1+\lambda^2}{2\lambda}\cosh(2\alpha)+\frac{1-\lambda^2}{2\lambda}\sinh(2\alpha)\cos(\omega t)$ and
$f^{(-)}=f^{(+)}(\alpha\rightarrow-\alpha)$.

Adding up all the harmonic oscillators on the lattice, we obtain the final expression of the complexity of TFD state
\begin{align}
\mathcal{C}_2=\frac{1}{2}\left\{\sum_{k_1,k_2=0}^{N-1}\sum_{\pm}\left[\log^2\left(f_{k_1,k_2}^{(\pm)}+\sqrt{(f_{k_1,k_2}^{(\pm)})^2-1}\right)
+\log^2\left(f^{(\pm)}+\sqrt{(f^{(\pm)})^2-1}\right)\right]\right\}^{1/2},\label{complexNTFD}
\end{align}
where
\begin{align}
f_{k_1,k_2}^{(\pm)}&=\frac{1}{2}\left(\frac{\mu}{\bar{\omega}_{k_1,k_2}}+\frac{\bar{\omega}_{k_1,k_2}}{\mu}\right)
\cosh(2\alpha_{k_1,k_2})\pm\frac{1}{2}\left(\frac{\mu}{\bar{\omega}_{k_1,k_2}}-\frac{\bar{\omega}_{k_1,k_2}}{\mu}\right)\cosh(2\alpha_{k_1,k_2})
\cos(\bar{\omega}_{k_1,k_2}t),\nonumber\\
f^{(\pm)}&=\frac{1}{2}\left(\frac{\mu}{\omega}+\frac{\omega}{\mu}\right)
\cosh(2\alpha)\pm\frac{1}{2}\left(\frac{\mu}{\omega}-\frac{\omega}{\mu}\right)\cosh(2\alpha)
\cos(\omega t),
\end{align}
and
\begin{align}
\alpha_{k_1,k_2}=\frac{1}{2}\log\left(\frac{1+e^{-\beta\bar{\omega}_{k_1,k_2}/2}}{1-e^{-\beta\bar{\omega}_{k_1,k_2}/2}}\right),\;\;\;\;
\alpha=\frac{1}{2}\log\left(\frac{1+e^{-\beta\omega/2}}{1-e^{-\beta\omega/2}}\right).
\end{align}
Note that $f_{k_1,k_2}^{(\pm)}$ and $f^{(\pm)}$ in (\ref{complexNTFD}) come respectively from the contributions of $\bar{x}_{k_1,k_2}$ and $\bar{y}_{k_1,k_2}$ in the Hamiltonian (\ref{Hdiag}).

\begin{figure}[h]
\begin{center}
\includegraphics[width=.32\textwidth]{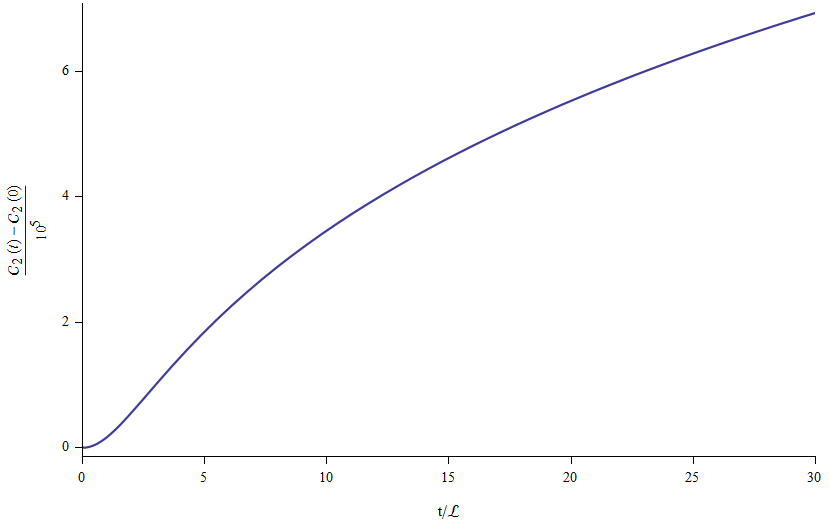}
\includegraphics[width=.32\textwidth]{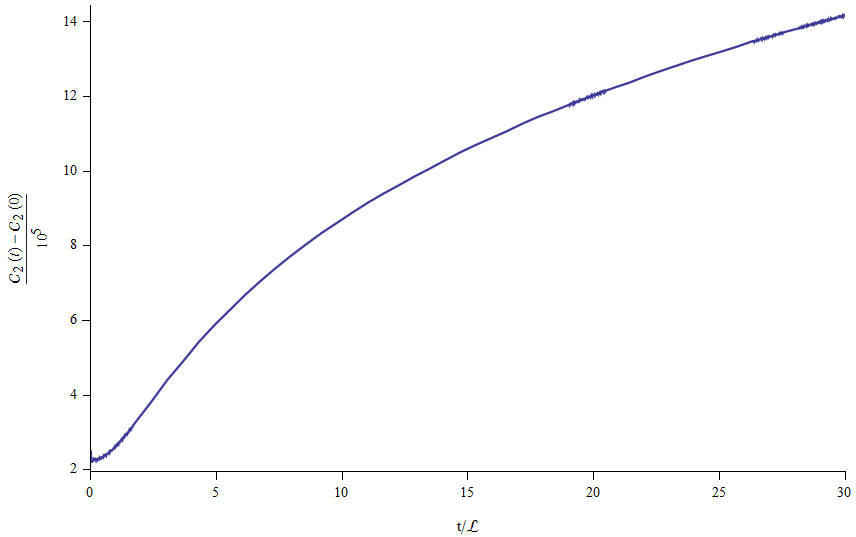}
\includegraphics[width=.32\textwidth]{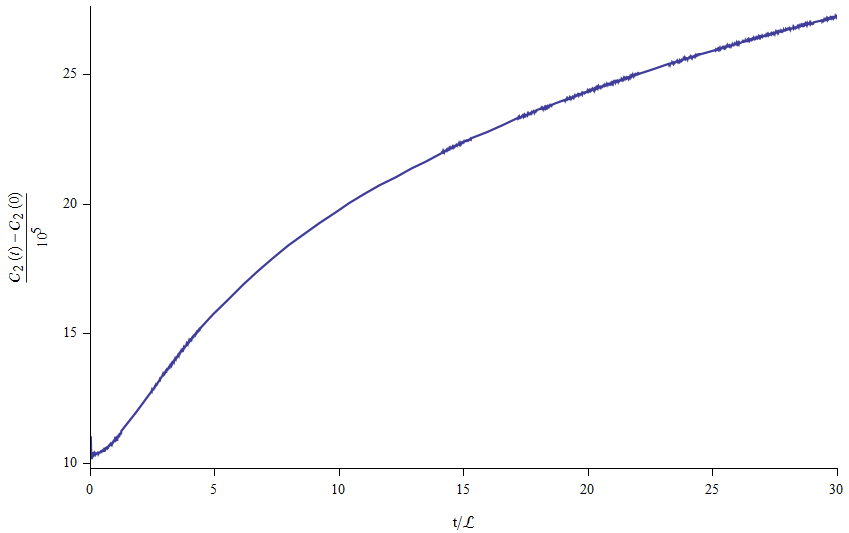}
\end{center}
\caption{Time dependence of complexity of TFD state. From left to right, the  plots correspond to $\beta=10^{-1}\mathcal{L}, 10^{-3}\mathcal{L}, 10^{-5}\mathcal{L}$ respectively. $\mathcal{L}$ is the size of the 2-lattice in each dimension. We insert 200 lattice sites in each dimension. Other parameters are fixed as $m=10^{-6}/\mathcal{L}, \mu=1/\mathcal{L}$.}
\label{fig1}
\end{figure}
Let's discuss the time dependence of the complexity of TFD state  (\ref{complexNTFD}).  In Fig.\ref{fig1}, we present the curves of complexities with different temperatures, the result is regularized by subtracting the initial value at $t=0$. One sees that all the curves fit the function $\sim\log^2(t)$ well. Saturation at high temperature found in Ref.\cite{1810.05151} is not found here, this is attributed to the contributions of $f^{(\pm)}$ and zero mode $f_{k_1=0,k_2=0}^{(\pm)}$ to complexity in (\ref{complexNTFD}) (Note that $f^{(\pm)}$ and $f_{k_1,k_2}^{(\pm)}$ correspond respectively to the $y$ and $x$ part of the Hamiltonian (\ref{Hdiag})). To see this we take the limit that  $m/\mu$, $\beta m$ and $mt$ are all small.  We take this limit due to we are interested in the small $m$ limit, where the QFT is close to CFT.  $m$ can't be zero since $m=0$ causes the singularity of complexity. In the limit, the contributions of $f^{\pm}$ and  zero mode  $f^{\pm}_{k_1=0,k_2=0}$ to complexity behave like
\begin{align}
\mathcal{C}_{2}\sim\frac{1}{4}\log^2\left(\frac{\mu t^2}{\beta}\right),
\end{align}
which explains the curves in Fig.\ref{fig1}.

\section{FS metric approach\label{section3}}
In this section, we study complexity in Proca theory with FS metric approach, we will consider both the ground state and the TFD state.
\subsection{complexity of ground state\label{CgroundFS}}
The FS line element is given by
\begin{align}
ds_{FS}(\sigma)=d\sigma\sqrt{|\partial_\sigma|\psi(\sigma)\rangle|^2-|\langle\psi(\sigma)|\partial_\sigma|\psi(\sigma)\rangle|^2}\label{FS},
\end{align}
which is used to measure the circuit length of a path going via states $|\psi(\sigma)\rangle$
\begin{align}
\ell(|\psi(\sigma)\rangle)=\int_{s_i}^{s_f}ds_{FS}(\sigma).\label{length}
\end{align}
The circuit we concerned is the one generated by the path-ordered unitary operator
\begin{align}
U(\sigma)=\overleftarrow{\mathcal{P}}e^{-i\int_{s_i}^\sigma G(s)ds}\label{FSU}
\end{align}
that transform a reference state to a  target state $|\psi\rangle_T=U|\psi\rangle_R$. Now the FS line element becomes
\begin{align}
ds_{FS}(\sigma)=d\sigma\sqrt{\langle G^2(\sigma)\rangle-\langle G(\sigma)\rangle^2},
\end{align}
which is independent of the path reparametrizations. Complexity is defined as the minimal length of the path
\begin{align}
\mathcal{C}=\mathop{\textrm{min}}\limits_{G(s)}\ell(|\psi(\sigma)\rangle)\label{FScom}
\end{align}
that connect the reference and target states.

In order to construct the desired target states, we first quantize Proca fields in terms of the creation and annihilation operators
\begin{align}
A_i(\vec{x})=\int\frac{d^2k}{\sqrt{2\omega_k}}\sum_{n=1}^2\varepsilon_i^{(n)}(\vec{k})[a^{(n)}_{\vec{k}}e^{-ik\cdot x}+
a^{(n)\dagger}_{\vec{k}}e^{ik\cdot x}],
\end{align}
where $\varepsilon_i^{(n)}$ are the polarization vectors. The commutation relation of creation and annihilation operators is given by $[a^{(i)}_{\vec{k}},a^{(j)\dagger}_{\vec{k}'}]=\delta_{i j}\delta(\vec{k}-\vec{k}')$, which preserves the commutation relation $[A_i(\vec{x}), \pi^j(\vec{x}')]=i\delta^j_i\delta(\vec{x}-\vec{x}')$.

The two-point correlation function in ground state takes the familiar form
\begin{align}
\langle 0|A^i(\vec{k})A^j(\vec{k}')|0\rangle=\frac{1}{2\omega_k}\delta^{ij}\delta(\vec{k}+\vec{k}'),
\end{align}
Now we introduce a reference state $|R(\nu)\rangle$ in such a way that the two-point function in this state takes the form
\begin{align}
\langle R(\nu)|A^i(\vec{k})A^j(\vec{k}')|R(\nu)\rangle=\frac{1}{2\nu}\delta^{ij}\delta(\vec{k}+\vec{k}').\label{2pointR}
\end{align}
similar to $\mu$ in (\ref{referenceN}), here the parameter $\nu$ is used to characterize the reference state. In coordinate space, one can see, from the two-point function $\langle R(\nu)|A^i(\vec{x})A^j(\vec{x}')|R(\nu)\rangle=\frac{1}{2\nu}\delta^{ij}\delta(\vec{x}-\vec{x}')$, that the fields  are unentangled in this state. Both the ground state and reference state are pure Gaussian states which are annihilated by the associated annihilation operators.

In order that, through the squeezing operator
\begin{align}
K(\vec{k})&=\sum_{i=1}^{2}\left(A_i(\vec{k})\pi_i(-\vec{k})+\pi_i(\vec{k})A_i(-\vec{k})\right),\nonumber\\
&=i\sum_{n=1}^2\left(a^{(n)\dagger}_{\vec{k}}a^{(n)\dagger}_{-\vec{k}}-a^{(n)}_{\vec{k}}a^{(n)}_{-\vec{k}}\right)
\end{align}
one can construct the approximate ground state
\begin{align}
|m^{(\Lambda)}\rangle=e^{-i\int_{k\leq\Lambda}d^2kr_k K(\vec{k})}|R(\nu)\rangle,\label{FSvac}
\end{align}
the creation and annihilation operators $b^{\dagger}_{\vec{k}}, b_{\vec{k}}$ associated to the reference state  are required to relate to the ones associated to the ground state through the Bogoliubov transformation
\begin{align}
b^{(n)}_{\vec{k}}=\beta^{+}_k a^{(n)}_{\vec{k}}+\beta^{-}_k a^{(n)\dagger}_{-\vec{k}},
\end{align}
with $\beta^{+}_k=\cosh(2r_k),\;\beta^{-}_k=\sinh(2r_k),\;r_k=\log\sqrt[4]{\frac{\nu}{\omega_k}}$. In terms of the creation and annihilation operators associated to  $|R(\nu)\rangle$, the squeezing operator is given by
\begin{align}
K(\vec{k})
=i\sum_{n=1}^2\left(b^{(n)\dagger}_{\vec{k}}b^{(n)\dagger}_{-\vec{k}}-b^{(n)}_{\vec{k}}b^{(n)}_{-\vec{k}}\right)
\end{align}
The state (\ref{FSvac}) is  identical to the ground state $|0\rangle$ up to a momentum cut-off $\Lambda$ and is usually used in cMERA\cite{1102.5524,1208.3469}. We choose this state as the target state in this subsection.

Now the the generator $G(\sigma)$ of circuit (\ref{FSU}) is given by
\begin{align}
G(\sigma)=\int_{k\leq\Lambda}d^2k K(\vec{k}) Y_{\vec{k}}(\sigma),\label{FSgenervacc}
\end{align}
with $Y_{\vec{k}}(s_f)$ being fixed to $r_k$  such that the circuit (\ref{FSU}) generated by (\ref{FSgenervacc}) connect the reference state $|R(\nu)\rangle$ and the target state $|m^{(\Lambda)}\rangle$.

Combining (\ref{FS}), (\ref{length}), (\ref{FSvac}) and (\ref{FSgenervacc}), one obtains complexity of the ground state
\begin{align}
\mathcal{C}=\mathop{\textrm{min}}\limits_{Y_{\vec{k}}(\sigma)}\int_{s_i}^{s_f}d\sigma \;2 \sqrt{\textrm{Vol}\int_{k\leq\Lambda}d^2k\left(\partial_\sigma Y_{\vec{k}}(\sigma)\right)^2}.\label{FScomK}
\end{align}
From (\ref{FScomK}) it is recognized that the space spanned by $Y_{\vec{k}}$ is Euclidean space. The shortest line in Euclidean space is straight line, which is parametrized as
\begin{align}
Y_{\vec{k}}(\sigma)=\frac{\sigma-s_i}{s_f-s_i}Y_{\vec{k}}(s_f).
\end{align}
Thus complexity of the ground state is given by
\begin{align}
\mathcal{C}=\sqrt{4\textrm{Vol}\int_{k\leq\Lambda}d^2k r^2_k}=\sqrt{\textrm{Vol}\int_{k\leq\Lambda}d^2k\frac{1}{4}\log^2\frac{\omega_k}{\nu}}.\label{complexGSK}
\end{align}
Note that this result is $\sqrt{2}$ times that of free scalar field theory, this is owning to the Proca fields have two spatial components in $2+1$-dimensional spacetime. It is interesting to note that ground state complexity (\ref{complexGSK}) share the similar form with that given in (\ref{complexNG}) which is obtained by Nielsen's approach.

Besides the squeezing operator $K(\vec{k})$,  one can also use the general quadratic operators
\begin{align}
K_{+}=\frac{1}{2}\sum_{n=1}^2b^{(n)\dagger}_{\vec{k}}b^{(n)\dagger}_{-\vec{k}},\;\;\;\;\;
K_{-}=\frac{1}{2}\sum_{n=1}^2b^{(n)}_{\vec{k}}b^{(n)}_{-\vec{k}},\;\;\;\;\;
K_{0}=\frac{1}{4}\sum_{n=1}^2b^{(n)\dagger}_{\vec{k}}b^{(n)}_{\vec{k}}+b^{(n)}_{-\vec{k}}b^{(n)\dagger}_{-\vec{k}}
\end{align}
which form the $\mathfrak{su}(1,1)$ algebra, to generate the approximate ground state
\begin{align}
|m^{(\Lambda)}\rangle\equiv e^{-i\frac{\pi}{4}\int_{k\leq\Lambda}d^2 k B(\vec{k},\nu)}|R(\nu)\rangle,\label{groundSU}
\end{align}
with
\begin{align}
B(\vec{k},\nu)=-2\sinh(2r_k)[K_{+}+K_{-}]+4\cosh(2r_k)K_0.
\end{align}
In order to calculate complexity of the state (\ref{groundSU}), let's consider the general state constructed with the $\mathfrak{su}(1,1)$ generators
\begin{align}
|\psi(\sigma)\rangle=e^{\int_{\Lambda}d^2k g(\vec{k},\sigma)}|R(\nu)\rangle,
\end{align}
where
\begin{align}
g(\vec{k},\sigma)=\alpha_{+}(\vec{k},\sigma)K_{+}(\vec{k})+\alpha_{-}(\vec{k},\sigma)K_{-}(\vec{k})
+\omega(\vec{k},\sigma)K_{0}(\vec{k}).\label{SUg}
\end{align}
The unitary generated by (\ref{SUg}) can be decomposed as\cite{Klimov}
\begin{align}
U(\sigma)=e^{\int_{\Lambda}d^2k \gamma_{+}(\vec{k},\sigma)K_{+}(\vec{k})}e^{\int_{\Lambda}d^2k \log\gamma_{0}(\vec{k},\sigma)K_{+}(\vec{k})}
e^{\int_{\Lambda}d^2k \gamma_{-}(\vec{k},\sigma)K_{-}(\vec{k})}.\label{decomp}
\end{align}
The coefficients are related through
\begin{align}
\gamma_{\pm}=\frac{2\alpha_{\pm}\sinh\Xi}{2\Xi\cosh\Xi-\omega\sinh\Xi},\;\;\gamma_{0}=(\cosh\Xi-\frac{\omega}{2\Xi}\sinh\Xi)^{-2},\;\;
\Xi^2=\frac{\omega^2}{4}-\alpha_{+}\alpha_{-}.\label{coeffrel}
\end{align}
After some operator algebra calculations, one obtains the following expression of complexity
\begin{align}
\mathcal{C}=\mathop{\textrm{min}}\limits_{\gamma_{+}(\vec{k},\sigma)}\int_{s_i}^{s_f}d\sigma\sqrt{\textrm{Vol}\int_{\Lambda}d^2k
\frac{\gamma'_{+}(\vec{k},\sigma)\gamma'^{*}_{+}(\vec{k},\sigma)}{(1-|\gamma_{+}(\vec{k},\sigma)|^2)^2}},\label{complexSU}
\end{align}
where $'$ denotes derivative respect to $\sigma$, and Vol is the volume of the time slice. From (\ref{complexSU})  it is recognized that for each momentum $\vec{k}$ the geometry is a Poincar\'{e} disk with complex coordinate $\gamma_{+}(\vec{k})$. The geodesic connecting the reference and ground states is a straight line on real axis. Therefore, the path generated by $B(\vec{k},\nu)$ is not the shortest one, since for $B(\vec{k},\nu)$
\begin{align}
\alpha_{\pm}(\vec{k},\sigma)&=\frac{i\pi}{2}\sinh(2r_k)\sigma,\;\;\;\;\;\omega(\vec{k},\sigma)=-i\pi\cosh(2r_k)\sigma,\;\;\nonumber\\
\gamma_{+}(\vec{k},\sigma)&=\frac{i\sinh(2r_k)\sin(\frac{\pi\sigma}{2})}{\cos(\frac{\pi\sigma}{2})+i\cosh(2r_k)\sin(\frac{\pi\sigma}{2})},
\end{align}
where $\gamma_{+}$ is not real. While the path generated by $K(\vec{k})$ coincides with the geodesic
\begin{align}
\alpha_{\pm}(\vec{k},\sigma)=\pm 2r_k\sigma,\;\;\;\;\;\;\;\;\omega(\vec{k},\sigma)=0,\;\;\;\;\;\;\;\;\gamma_{+}(\vec{k},\sigma)=\tanh(2r_k\sigma).
\end{align}
Thus with the more general $\mathfrak{su}(1,1)$ generators, one still gets the result that the straight line corresponds to the shortest path, the complexity is still given by Eq.(\ref{complexGSK}).

\subsection{complexity of \textrm{TFD} state\label{CTFDFS}}
At $t=0$, the TFD state is given by
\begin{align}
|\textrm{TFD}(0)\rangle&=N e^{\int d^2k \;e^{-\beta \omega_k/2} (a_{\vec{k}}^{(1)L\dagger}a_{\vec{k}}^{(1)R\dagger}+a_{\vec{k}}^{(2)L\dagger}a_{\vec{k}}^{(2)R\dagger})}|0\rangle\nonumber\\
&=N e^{\int d^2k \;e^{-\beta \omega_k/2} (a_{\vec{k}}^{(1)L\dagger}a_{\vec{k}}^{(1)R\dagger}+a_{\vec{k}}^{(2)L\dagger}a_{\vec{k}}^{(2)R\dagger})}|0\rangle.\label{FSTFD0}
\end{align}
This state  is annihilated by operators $c_{\vec{k}}^{(n)}, \tilde{c}_{-\vec{k}}^{(n)}$ which relate to creation and annihilation operators of the vacuum state $|0\rangle$ through Bogoliubov transformations
\begin{align}
c^{(n)}_{\vec{k}}=\cosh\alpha_k a_{\vec{k}}^{(n)L}-\sinh\alpha_k a_{-\vec{k}}^{(n)R\dagger},\;\;\;\;\;\;\;\;
\tilde{c}^{(n)}_{-\vec{k}}=\cosh\alpha_k a_{-\vec{k}}^{(n)R}-\sinh\alpha_k a_{\vec{k}}^{(n)L\dagger},
\end{align}
where $\tanh\alpha_k=e^{-\beta\omega_k/2}$.

The time-dependent TFD state can be achieved by action of the time evolution operator on the $t=0$ TFD state
\begin{align}
|\textrm{TFD}(t)\rangle=e^{-iHt}|\textrm{TFD}(0)\rangle,
\end{align}
where $t_L=t_R=t$ has been used, and
\begin{align}
H=\frac{1}{2}\int d^2k\left[\omega_k\sum_{n=1}^2\left(a_{\vec{k}}^{(n)L\dagger}a_{\vec{k}}^{(n)L}+a_{\vec{k}}^{(n)R\dagger}a_{\vec{k}}^{(n)R}+1\right)\right]
\end{align}
is the Hamiltonian of the whole system.

The general quadratic operators which form $\mathfrak{su}(1,1)$ algebra are defined as
\begin{align}
K_{+}=\frac{1}{2}\sum_{n=1}^2(c^{(n)\dagger}_{\vec{k}}\tilde{c}^{(n)\dagger}_{-\vec{k}}),\;\;\;
K_{-}=\frac{1}{2}\sum_{n=1}^2(c^{(n)}_{\vec{k}}\tilde{c}^{(n)}_{-\vec{k}}),\;\;\;
K_{0}=\frac{1}{4}\sum_{n=1}^2(c^{(n)\dagger}_{\vec{k}}c^{(n)}_{\vec{k}}+\tilde{c}^{(n)}_{-\vec{k}}\tilde{c}^{(n)\dagger}_{-\vec{k}}),
\end{align}
with which the time-dependent TFD state can be rewritten as
\begin{align}
|\textrm{TFD}(t)\rangle=e^{-i\frac{1}{2}\int d^2k\;\omega_k  t[4\cosh(2\alpha_k) K_0+2\sinh(2\alpha_k)(K_{+}+K_{-})]}|\textrm{TFD}(0)\rangle.\label{FSTFDt1}
\end{align}
The coefficients can be read out through comparing (\ref{FSTFDt1}) with (\ref{SUg})
\begin{align}
\alpha_{\pm}=-i\omega_k t\sinh(2\alpha_k),\;\;\;\;\;\;\;\;\;\;\omega=-2i\omega_k t\cosh(2\alpha_k).
\end{align}
Thus the unitary in (\ref{FSTFDt1}) can be decomposed as
\begin{align}
|\textrm{TFD}(t)\rangle=e^{\int d^2k \gamma_{+}K_{+}(\vec{k})}
e^{\int d^2k \log\gamma_{0}K_{0}(\vec{k})} e^{\int d^2k \gamma_{-}K_{-}(\vec{k})}|\textrm{TFD}(0)\rangle,\label{FSTFDt2}
\end{align}
with
\begin{align}
\gamma_{\pm}=\frac{-i\sinh(2\alpha_k)\sin\Xi}{\cos\Xi+i\cosh(2\alpha_k)\sin\Xi},\;\;\;\;\;\;\;\;\Xi=\omega_k t.
\end{align}
In terms of the path parameter $\sigma$, $\gamma_{+}$ can be written as
\begin{align}
\gamma_{\pm}(\vec{k}, \sigma)=\frac{-i\sinh(2\alpha_k)\sin(\omega_k t \sigma)}{\cos(\omega_k t \sigma)+i\cosh(2\alpha_k)\sin(\omega_k t \sigma)}.
\end{align}

We choose $|\textrm{TFD}(0)\rangle$  as the reference state and $|\textrm{TFD}(t)\rangle$ as the target state, then the complexity is given by
\begin{align}
\mathcal{C}&=\mathop{\textrm{min}}\limits_{\gamma_{+}(\vec{k},\sigma)}\int_{s_i}^{s_f}d\sigma\sqrt{\textrm{Vol}\int d^2k
\frac{\gamma'_{+}(\vec{k},\sigma)\gamma'^{*}_{+}(\vec{k},\sigma)}{(1-|\gamma_{+}(\vec{k},\sigma)|^2)^2}}\nonumber\\
&=\int_{s_i}^{s_f}d\sigma\sqrt{\textrm{Vol}\int d^2k\;
\omega_k^2 t^2\sinh^2(2\alpha_k)}\nonumber\\
&=\frac{2t}{\beta^2} \sqrt{\textrm{Vol}\left[6\textrm{Li}_3(e^{-m\beta})+6\beta m\textrm{Li}_2(e^{-m\beta})+3\beta^2m^2\textrm{Li}_1(e^{-m\beta})+\beta^3 m^3\textrm{Li}_0(e^{-m\beta})\right]},
\label{complexFSt1}
\end{align}
where $\textrm{Li}_n(\cdot)$ is the polylog function, and in the last equality $\beta$ is assumed to be large, i.e., we consider the low temperature limit. For $m=0$, we have
\begin{align}
\mathcal{C}
&=\frac{2t}{\beta^2} \sqrt{\textrm{Vol} \;6\zeta(3)},
\label{complexFSt1m0}
\end{align}
where $\zeta(\cdot)$ is the Riemann zeta function. The results (\ref{complexFSt1}) and (\ref{complexFSt1m0}) show that complexity grows linearly with time, which agree with the late-time growth behavior of gravitational action evaluated on WDW patch.

\section{Summary}
In this paper, we study complexities of the states in Proca theory with Nielsen's approach and FS metric approach. For both the approaches, we all consider the ground and TFD states. To study complexity with Nielsen's approach, we regularize the theory by placing the fields on a lattice, and obtain a system of coupled harmonic oscillators. Through choosing proper coordinates, we recast the system to the one of decoupled harmonic oscillators, based on which we give the ground state of the system.  With the aid of the covariance matrix approach, we give complexities of the ground and TFD states. We compare our results with that of the free scalar field theory, for both the ground state and the TFD state one finds extra contributions, which correspond to the $y$ part in the  Hamiltonian of decoupled harmonic oscillators (\ref{Hdiag}), to complexity.  We examine the time dependence of TFD-state complexity and find it exhibits a logarithmic growth $\log^2t$ while the saturation at high temperature found in free scalar field theory is not found in our case. Analysis shows that the growth behavior of TFD-state complexity is due to the contributions of both $f^{\pm}$  and the zero mode $f_{k_1=0,k_2=0}^{(\pm)}$ which correspond to the $y$ and $x$ parts in (\ref{Hdiag}) respectively, while for free scalar field theory only the zero mode $f_{k_1=0,k_2=0}^{(\pm)}$ is responsible for the logarithmic growth of complexity.

To study complexity with FS metric approach,  first we give the approximate ground state with the squeezing operator and calculate circuit length with FS metric. Calculations show that the coefficients of the squeezing operator form  Euclidean space, which implies that the straight line corresponds to complexity of the state. Since the spatial components of the Proca fields is two in $2+1$ dimensions, complexity of the ground state we obtained is $\sqrt{2}$ times that of free scalar field. Then we give the approximate ground state with the general $\mathfrak{su}(1,1)$ generators, in this case still the straight line give the complexity, which is just the one given by the squeezing operator. We choose the TFD state at $t=0$ as the reference state and the TFD state at arbitrary $t$ as the target state, and calculate the circuit length with FS metric. The result complexity exhibits the linear dependence of time, which agrees with the late-time behavior of holographic complexity.

For further directions, it would be interesting to discuss other cost functions $F_1, F_p$ and $F_q$ for Nielsen's approach. Furthermore, it would be interesting to study complexities in charged vector field theory and Yang-Mills theory, and discuss the effects of gauge symmetries on complexity.

\section*{Acknowledgment}
The work of LZC is supported by the National Natural Science Foundation of China (No.62005199), and the Natural Science Foundation of Shandong Province (Nos. ZR2020 LLZ001 and ZR2019LLZ006). The work of YY is supported by the Key Research and Development Plan of Shandong Province (No. 2019GGX101073). The work of JQZ is supported by Natural Science Foundation of Shandong Province (No. ZR2020KF017).

\providecommand{\href}[2]{#2}\begingroup
\footnotesize\itemsep=0pt
\providecommand{\eprint}[2][]{\href{http://arxiv.org/abs/#2}{arXiv:#2}}


\begin{thebibliography}{}

\bibitem{0603001}
S.~Ryu and T.~Takayanagi,
``Holographic derivation of entanglement entropy from AdS/CFT,''
Phys. Rev. Lett. \textbf{96}, 181602 (2006)
[arXiv:hep-th/0603001 [hep-th]].

\bibitem{1306.0533}
J.~Maldacena and L.~Susskind,
``Cool horizons for entangled black holes,''
Fortsch. Phys. \textbf{61}, 781-811 (2013)
[arXiv:1306.0533 [hep-th]].


\bibitem{1411.0690}
L.~Susskind,
``Entanglement is not enough,''
Fortsch. Phys. \textbf{64}, 49-71 (2016)
[arXiv:1411.0690 [hep-th]].


\bibitem{1406.2678}
D.~Stanford and L.~Susskind,
``Complexity and Shock Wave Geometries,''
Phys. Rev. D \textbf{90}, no.12, 126007 (2014)
[arXiv:1406.2678 [hep-th]].


\bibitem{1408.2823}
L.~Susskind and Y.~Zhao,
``Switchbacks and the Bridge to Nowhere,''
[arXiv:1408.2823 [hep-th]].


\bibitem{1509.07876}
A.~R.~Brown, D.~A.~Roberts, L.~Susskind, B.~Swingle and Y.~Zhao,
``Holographic Complexity Equals Bulk Action?,''
Phys. Rev. Lett. \textbf{116}, no.19, 191301 (2016)
[arXiv:1509.07876 [hep-th]].


\bibitem{1512.04993}
A.~R.~Brown, D.~A.~Roberts, L.~Susskind, B.~Swingle and Y.~Zhao,
``Complexity, action, and black holes,''
Phys. Rev. D \textbf{93}, no.8, 086006 (2016)
[arXiv:1512.04993 [hep-th]].


\bibitem{1703.06297}
J.~Tao, P.~Wang and H.~Yang,
``Testing holographic conjectures of complexity with Born\textendash{}Infeld black holes,''
Eur. Phys. J. C \textbf{77}, no.12, 817 (2017)
[arXiv:1703.06297 [hep-th]].


\bibitem{1612.03627}
W.~J.~Pan and Y.~C.~Huang,
``Holographic complexity and action growth in massive gravities,''
Phys. Rev. D \textbf{95}, no.12, 126013 (2017)
[arXiv:1612.03627 [hep-th]].


\bibitem{1606.08307}
R.~G.~Cai, S.~M.~Ruan, S.~J.~Wang, R.~Q.~Yang and R.~H.~Peng,
``Action growth for AdS black holes,''
JHEP \textbf{09}, 161 (2016)
[arXiv:1606.08307 [gr-qc]].


\bibitem{1806.06216}
R.~Auzzi, S.~Baiguera, M.~Grassi, G.~Nardelli and N.~Zenoni,
``Complexity and action for warped AdS black holes,''
JHEP \textbf{09}, 013 (2018)
[arXiv:1806.06216 [hep-th]].

\bibitem{1804.07410}
S.~Chapman, H.~Marrochio and R.~C.~Myers,
``Holographic complexity in Vaidya spacetimes. Part I,''
JHEP \textbf{06}, 046 (2018)
[arXiv:1804.07410 [hep-th]].


\bibitem{1805.07262}
S.~Chapman, H.~Marrochio and R.~C.~Myers,
``Holographic complexity in Vaidya spacetimes. Part II,''
JHEP \textbf{06}, 114 (2018)
[arXiv:1805.07262 [hep-th]].


\bibitem{1806.10312}
J.~Jiang and H.~Zhang,
``Surface term, corner term, and action growth in $F(R_{abcd})$ gravity theory,''
Phys. Rev. D \textbf{99}, no.8, 086005 (2019)
[arXiv:1806.10312 [hep-th]].


\bibitem{1808.09917}
S.~Mahapatra and P.~Roy,
``On the time dependence of holographic complexity in a dynamical Einstein-dilaton model,''
JHEP \textbf{11}, 138 (2018)
[arXiv:1808.09917 [hep-th]].


\bibitem{1808.00067}
S.~A.~Hosseini Mansoori, V.~Jahnke, M.~M.~Qaemmaqami and Y.~D.~Olivas,
``Holographic complexity of anisotropic black branes,''
Phys. Rev. D \textbf{100}, no.4, 046014 (2019)
[arXiv:1808.00067 [hep-th]].


\bibitem{2107.08608}
C.~Bai, W.~H.~Li and X.~H.~Ge,
``Towards the non-equilibrium thermodynamics of the complexity and the Jarzynski identity,''
[arXiv:2107.08608 [hep-th]].

\bibitem{1810.02208}
K.~Meng,
``Holographic complexity of Born\textendash{}Infeld black holes,''
Eur. Phys. J. C \textbf{79}, no.12, 984 (2019)
[arXiv:1810.02208 [hep-th]].


\bibitem{0502070}
M.~Nielsen,
``A geometric approach to quantum circuit lower bounds,''
[arXiv:quant-ph/0502070 [quant-ph]].


\bibitem{1707.08570}
R.~Jefferson and R.~C.~Myers,
``Circuit complexity in quantum field theory,''
JHEP \textbf{10}, 107 (2017)
[arXiv:1707.08570 [hep-th]].


\bibitem{1801.07620}
R.~Khan, C.~Krishnan and S.~Sharma,
``Circuit Complexity in Fermionic Field Theory,''
Phys. Rev. D \textbf{98}, no.12, 126001 (2018)
[arXiv:1801.07620 [hep-th]].


\bibitem{1803.10638}
L.~Hackl and R.~C.~Myers,
``Circuit complexity for free fermions,''
JHEP \textbf{07}, 139 (2018)
[arXiv:1803.10638 [hep-th]].


\bibitem{1810.05151}
S.~Chapman, J.~Eisert, L.~Hackl, M.~P.~Heller, R.~Jefferson, H.~Marrochio and R.~C.~Myers,
``Complexity and entanglement for thermofield double states,''
SciPost Phys. \textbf{6}, no.3, 034 (2019)
[arXiv:1810.05151 [hep-th]].


\bibitem{1812.00193}
J.~Jiang and X.~Liu,
``Circuit Complexity for Fermionic Thermofield Double states,''
Phys. Rev. D \textbf{99}, no.2, 026011 (2019)
[arXiv:1812.00193 [hep-th]].


\bibitem{1910.08806}
M.~Doroudiani, A.~Naseh and R.~Pirmoradian,
``Complexity for Charged Thermofield Double States,''
JHEP \textbf{01}, 120 (2020)
[arXiv:1910.08806 [hep-th]].


\bibitem{1909.10557}
E.~Caceres, S.~Chapman, J.~D.~Couch, J.~P.~Hernandez, R.~C.~Myers and S.~M.~Ruan,
``Complexity of Mixed States in QFT and Holography,''
JHEP \textbf{03}, 012 (2020)
[arXiv:1909.10557 [hep-th]].

\bibitem{2004.00344}
M.~Guo, Z.~Y.~Fan, J.~Jiang, X.~Liu and B.~Chen,
``Circuit complexity for generalized coherent states in thermal field dynamics,''
Phys. Rev. D \textbf{101}, no.12, 126007 (2020)
[arXiv:2004.00344 [hep-th]].



\bibitem{1907.08223}
A.~Bhattacharyya, P.~Nandy and A.~Sinha,
``Renormalized Circuit Complexity,''
Phys. Rev. Lett. \textbf{124}, no.10, 101602 (2020)
[arXiv:1907.08223 [hep-th]].

\bibitem{1707.08582}
S.~Chapman, M.~P.~Heller, H.~Marrochio and F.~Pastawski,
``Toward a Definition of Complexity for Quantum Field Theory States,''
Phys. Rev. Lett. \textbf{120}, no.12, 121602 (2018)
[arXiv:1707.08582 [hep-th]].


\bibitem{1710.00600}
R.~Q.~Yang, C.~Niu, C.~Y.~Zhang and K.~Y.~Kim,
``Comparison of holographic and field theoretic complexities for time dependent thermofield double states,''
JHEP \textbf{02}, 082 (2018)
[arXiv:1710.00600 [hep-th]].

\bibitem{1902.01912}
M.~Sinamuli and R.~B.~Mann,
``Holographic Complexity and Charged Scalar Fields,''
Phys. Rev. D \textbf{99}, no.10, 106013 (2019)
[arXiv:1902.01912 [hep-th]].

\bibitem{2103.06920}
N.~Chagnet, S.~Chapman, J.~de Boer and C.~Zukowski,
``Complexity for Conformal Field Theories in General Dimensions,''
[arXiv:2103.06920 [hep-th]].


\bibitem{1703.00456}
P.~Caputa, N.~Kundu, M.~Miyaji, T.~Takayanagi and K.~Watanabe,
``Anti-de Sitter Space from Optimization of Path Integrals in Conformal Field Theories,''
Phys. Rev. Lett. \textbf{119}, no.7, 071602 (2017)
[arXiv:1703.00456 [hep-th]].

\bibitem{1706.07056}
P.~Caputa, N.~Kundu, M.~Miyaji, T.~Takayanagi and K.~Watanabe,
``Liouville Action as Path-Integral Complexity: From Continuous Tensor Networks to AdS/CFT,''
JHEP \textbf{11}, 097 (2017)
[arXiv:1706.07056 [hep-th]].

\bibitem{2011.08188}
J.~Boruch, P.~Caputa and T.~Takayanagi,
``Path-Integral Optimization from Hartle-Hawking Wave Function,''
Phys. Rev. D \textbf{103}, no.4, 046017 (2021)
[arXiv:2011.08188 [hep-th]].

\bibitem{1807.04422}
P.~Caputa and J.~M.~Magan,
``Quantum Computation as Gravity,''
Phys. Rev. Lett. \textbf{122}, no.23, 231302 (2019)
[arXiv:1807.04422 [hep-th]].




\bibitem{1811.05985}
T.~Ali, A.~Bhattacharyya, S.~Shajidul Haque, E.~H.~Kim and N.~Moynihan,
``Post-Quench Evolution of Complexity and Entanglement in a Topological System,''
Phys. Lett. B \textbf{811}, 135919 (2020)
[arXiv:1811.05985 [hep-th]].



\bibitem{1902.10720}
F.~Liu, S.~Whitsitt, J.~B.~Curtis, R.~Lundgren, P.~Titum, Z.~C.~Yang, J.~R.~Garrison and A.~V.~Gorshkov,
``Circuit complexity across a topological phase transition,''
Phys. Rev. Res. \textbf{2}, no.1, 013323 (2020)
[arXiv:1902.10720 [quant-ph]].


\bibitem{1906.11279}
Z.~Xiong, D.~X.~Yao and Z.~Yan,
``Nonanalyticity of circuit complexity across topological phase transitions,''
Phys. Rev. B \textbf{101}, no.17, 174305 (2020)
[arXiv:1906.11279 [cond-mat.str-el]].

\bibitem{2106.12648}
U.~Sood and M.~Kruczenski,
``Circuit complexity near critical points,''
[arXiv:2106.12648 [quant-ph]].


\bibitem{1808.03105}
A.~Bhattacharyya, A.~Shekar and A.~Sinha,
``Circuit complexity in interacting QFTs and RG flows,''
JHEP \textbf{10}, 140 (2018)
[arXiv:1808.03105 [hep-th]].

\bibitem{2109.09759}
K.~Adhikari, S.~Choudhury, S.~Kumar, S.~Mandal, N.~Pandey, A.~Roy, S.~Sarkar, P.~Sarker and S.~S.~Shariff,
``Circuit Complexity in $\mathcal{Z}_{2}$${\cal EEFT}$,''
[arXiv:2109.09759 [hep-th]].


\bibitem{2108.08208}
A.~Moghimnejad and S.~Parvizi,
``Circuit Complexity in $U(1)$ Gauge Theory,''
[arXiv:2108.08208 [hep-th]].

\bibitem{1102.5524}
J.~Haegeman, T.~J.~Osborne, H.~Verschelde and F.~Verstraete,
``Entanglement Renormalization for Quantum Fields in Real Space,''
Phys. Rev. Lett. \textbf{110}, no.10, 100402 (2013)
[arXiv:1102.5524 [hep-th]].

\bibitem{1208.3469}
M.~Nozaki, S.~Ryu and T.~Takayanagi,
``Holographic Geometry of Entanglement Renormalization in Quantum Field Theories,''
JHEP \textbf{10}, 193 (2012)
[arXiv:1208.3469 [hep-th]].


\bibitem{Klimov}
A.~Klimov and S.~Chumakov,
$A \;group$-$theoretical \;approach\; to \;quantum\; optics: models \;of \;atom$-$field \;interactions$
(Wiley-VCH, 2009).


\end{thebibliography}
\end{document}